\def\breakon{\end{multicols}\widetext\vspace{-.2cm}
\noindent\rule{.48\linewidth}{.3mm}\rule{.3mm}{.3cm}\vspace{.0cm}}
\def\breakoff{\vspace{-.2cm}
\noindent
\rule{.52\linewidth}{.0mm}\rule[-.27cm]{.3mm}{.3cm}\rule{.48\linewidth}{.3mm}
\vspace{-.3cm}
\begin{multicols}{2}
\narrowtext}
\documentstyle[aps,prl,multicol,epsfig]{revtex}

\newcommand{\bea}{\begin{eqnarray}}
\newcommand{\eea}{\end{eqnarray}}
\newcommand{\non}{\nonumber}
\newcommand{\w}{\tilde}
\begin{document}

\draft

\widetext

\title{Geometric frustration and  magnetization plateaus in
quantum spin and Bose-Hubbard models on tubes} 

\author{Dmitry Green$^a$ and Claudio Chamon$^b$}
\address{$^a$ Department of Physics, Yale University,
New Haven, CT 06520\\ $^b$ Department of Physics, Boston University,
Boston, MA 02215}

\maketitle

\begin{abstract}
We study XXZ Heisenberg models on frustrated triangular lattices wrapped
around a cylinder. In addition to having interesting magnetic phases, these
models are equivalent to Bose-Hubbard models that describe the physical
problem of adsorption of noble gases on the surface of carbon nanotubes. We
find analytical results for the possible magnetization plateau values as a
function of the wrapping vectors of the cylinder, which in general introduce
extra geometric frustration besides the one due to the underlying triangular
lattice. We show that for particular wrapping vectors $(N,0)$, which
correspond to the zig-zag nanotubes, there is a macroscopically degenerate
ground state in the classical Ising limit. The Hilbert space for the
degenerate states can be enumerated by a mapping first into a path in a
square lattice wrapped around a cylinder (a Bratteli diagram), and then to free fermions interacting with a single ${\bf
Z}_N$ degree of freedom. From this model we obtain the spectrum in the
anisotropic Heisenberg limit, showing that it is gapless.  The continuum
limit is a $c=1$ conformal field theory with compactification radius $R=N$
set by the physical tube radius. This result cannot be checked against a
Lieb-Schultz-Mattis argument, for the argument is inconclusive when applied
to this problem. We show that the compactification radius quantization is
exact in the projective $J_\perp/J_z \ll 1$ limit, and that higher order corrections reduce the value of $R$. The particular case of a $(N=2,0)$
tube, which corresponds to a 2-leg ladder with cross links, is studied
separately and shown to be gapped because the fermion mapped problem contains
superconducting pairing terms.
\end{abstract}

\begin{multicols}{2}

\narrowtext

\section{Introduction}

There has been growing interest in one-dimensional quantum spin
chains and ladders that display magnetization plateaus when subjected to an
external magnetic field
\cite{Hida,Okamoto,Oshikawa,Cabra1,Cabra2,Andrei1,Andrei2}. This requires
materials with exchange couplings within a range for which the necessary
external fields are attainable in the laboratory. Examples are the spin $S=1$
nickel compound
[Ni(Medpt)$_2$($\mu$-ox)($\mu$--N$_3$)]ClO$_4$$\cdot$0.5H$_2$O, which
displays a plateau at $\langle M\rangle =1/2$ \cite{Narumi}, and the spin
$S=1/2$ compound NH$_4$CuCl$_3$, which displays plateaus with magnetization
$\langle M\rangle =1/4,3/4$ \cite{Shiramura} (the magnetization is measured
as a fraction of the maximum magnetization {\it per} spin $S$). One of the
most interesting features in these systems is the possibility of gapless
plateaus even in integer spin systems
\cite{Oshikawa}.

An important ingredient to obtain the plateaus is a $p$-merization, or the
presence of periodic structures that allow for ground states with non-zero
$\langle M\rangle$ other than the trivial cases $\langle M\rangle=\pm 1$. A
particular case of a $p$-merized structure is a ``spin tube'', like the one
proposed theoretically by Cabra, Honecker, and Pujol \cite{Cabra2}, and by
Citro, Orignac, Andrei, and co-workers \cite{Andrei1,Andrei2}. These tubes
basically consist of $p$-leg spin ladders with periodic boundary conditions
coupling the 1st and $p$th chains. Although interesting structures from a
theoretical perspective, these tubes are not yet realizable experimentally.

In this paper we explore another type of spin tube, which is formed by
wrapping a triangular lattice on a cylinder. The triangular lattice with
anti-ferromagnetic couplings is frustrated, and, depending on the wrapping
vector, extra geometric frustrations are introduced. One motivation to study
such types of spin tube lattices is that they are realized physically in
monolayer adsorption of noble gases on the surface of carbon nanotubes. This
paper contains a detailed account of the results in Ref.~\cite{Green-Chamon},
as well as new analytical results that explain the numerical findings of the
previous work.
\begin{figure}
\noindent
\center
\epsfxsize=2.2 in
\epsfbox{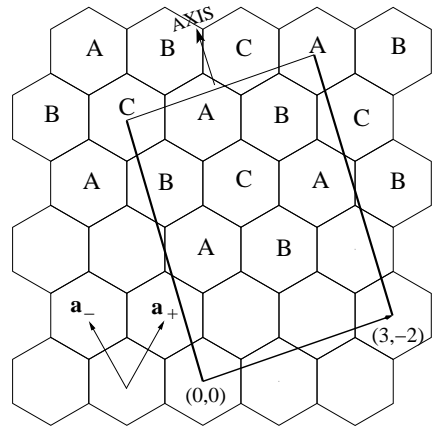}
\vspace{.05 in}
\caption{An example of wrapping of the graphite sheet to make a $(3,-2)$
tube. ${\vec a}_\pm$ are the primitive lattice vectors of the honeycomb
lattice.  The solid rectangle is the supercell of the tube.  Also shown is the tripartite labeling A, B, C.}
\label{fig:TubeCell}
\end{figure}

The lattice sites for the spin tubes we consider can be described by a pair of
integers $(N,M)$, exactly the notation for carbon nanotubes, for which the
following nomenclature applies: $(N,N)$ are called armchairs, $(N,0)$ are
zig-zags, and the general case are chiral. The triangular lattice points are
the centers of the hexagonal lattice of the nanotubes (the adsorption
centers), as shown in Fig. \ref{fig:TubeCell}. The integers $(N,M)$ define the wrapping vector of the honeycomb lattice that identifies the origin with the point $N {\vec a_+}+ M{\vec a_-}\equiv 0$, where $\vec a_\pm$ are the primitive lattice vectors of the honeycomb lattice.

The tripartite nature of the triangular lattice of adsorption sites is
destroyed whenever the wrapping vector $(N,M)$ is such that $N-M$ is not
divisible by $3$. An example of a $(7,0)$ tube is presented in
Ref.~\cite{Green-Chamon}, and reproduced below in Fig.~\ref{fig:Tube}.
\begin{figure}
\noindent
\center
\vspace{-.6 in}
\epsfxsize=3.0 in
\epsfbox{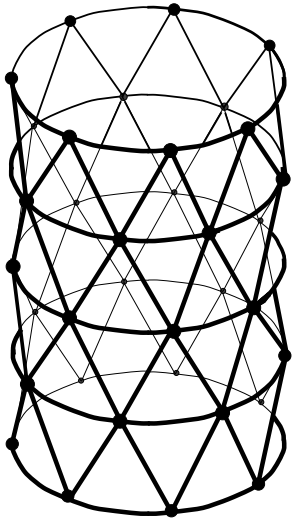}
\vspace{-.5 in}
\caption{Adsorption sites on a $(7,0)$ zig-zag nanotube}
\label{fig:Tube}
\end{figure}
The problem of adsorption of noble gases onto planar graphite can be
understood within a lattice gas model, with the hexagonal substrate providing
adsorption sites that form a triangular lattice \cite{Schick}. This problem
is equivalent to a Bose-Hubbard model, where a large nearest neighbor
repulsion arises from the fact that even the smaller of noble gas atoms,
helium, spread over an area larger than that of the carbon hexagons (the
characteristic length scale for the zero-point motion of the He atom comes
from the Lennard-Jones potential length scale $\sigma$=2.56\AA, while the
separation between C atoms in the hexagons is 1.42\AA) \cite{Bretz}. The
potential energy for atoms to be adsorbed in the same site are even larger,
and thus we take the hard-core boson limit. The case of adsorption on the
nanotube geometry is similar once the periodicity condition due to the
wrapping vector $(N,M)$ is considered.

The lattice gas is defined by the Bose-Hubbard
Hamiltonian\cite{Murthy,Auerbach}
\begin{equation}
 \label{eq:Hubbard}
{\cal H}\! =\! -t\sum_{\langle ij\rangle} (b_i^\dagger b_j + b_j^\dagger
 b_i)
 + V\sum_{\langle ij\rangle} n_i n_j - \mu\sum_i n_i~,
\end{equation}
where $n_i$ is the boson density at site $i$, $V$ is the nearest neighbor
repulsion and $t$ is the hopping amplitude. In the equivalent Heisenberg spin
representation,
\begin{equation}
 \label{eq:Spin}
\!\!\!\!{\cal H}\! =\!  -J_\perp  
\sum_{\langle ij\rangle} \left(S^x_i S^x_j\! +\! S^y_i S^y_j \right)
 \!-\!J_z \sum_{\langle ij\rangle} S^z_i S^z_j \!-\! H\sum_i S^z_i
\end{equation}
where $S^z_i=n_i-1/2$, $J_\perp=2t$, $J_z=-V$, and $H=\mu-3V$ is an effective
external magnetic field. The equivalence between the hard core Bose-Hubbard
model and the XXZ Heisenberg spin $S=1/2$ model allows us to study both
physical problems of the magnetization properties of the spin tube and the
adsorption of noble gases on carbon nanotubes at the same time.

In the adsorption problem one finds filling fraction plateaus as a function
of the chemical potential; this is the counterpart of the magnetization
plateau in the spin models as a function of external magnetic field. The
filling fraction plateaus most often correspond to solid phases, with an
energy gap in the excitation spectrum. However, as we show in this paper,
there are examples of gapless plateaus which occur in the case of $(N,0)$
tubes (zig-zags) with $N$ not divisible by 3. These plateaus correspond to a
compressible correlated fluid, which is described by a $c=1$ conformal field
theory with compactification radius $R=N$.

Many of the interesting results we find in the study of the spin tubes come
from the geometric frustration that is introduced when the lattice is wrapped
up into a cylinder. In some cases, one can think of the boundary conditions
as forcing a single domain wall running along the tube. Recently, Henley and
Zhang \cite{Henley} have studied the nearest neighbor Hubbard model for
spinless fermions as a toy model for understandying stripe phases when the
system is doped away from half-filling. The situation is very similar to our
case.

The paper is organized as follows. In section \ref{sec:classical} we study
the classical limit when the hopping $t$ in the Bose-Hubbard model goes to
zero; in spin language, this corresponds to the Ising limit. We explain a
macroscopic degeneracy present in the frustrated $(N,0)$ zig-zag tubes. In section
\ref{sec:quantum} we turn on the quantum hopping term $t$ (or the XY spin
couplings $J_\perp=2t$). We sumarize our findings in section
\ref{sec:conclusion}, where we also discuss possible experimental signatures
of the plateaus.

%
%
\section{Classical limit}
\label{sec:classical}

We start our analysis in the simpler classical limit, which corresponds to
the Ising limit $J_\perp=0$. In the Bose-Hubbard model it is the no-hopping
limit $t=0$. We will carry out the discussion interchangeably between the
Bose-Hubbard and spin languages.

Without loss of generality, we take the integers $N,M>0$; all other cases can always be brought to this
form by a rotation of the basis vectors ${\vec a_\pm}$ by multiples of
60$^\circ$. The triangular lattice sites can be partitioned into three
interpenetrating sublattices that we label by $A,B$, and $C$.  If we define
$q\equiv(N-M)(\rm mod\;3)$, then the wrapping of the lattice does not break
this tripartite property if $q=0$.

\begin{figure}
\noindent
\center
\epsfxsize=2.0 in
\epsfbox{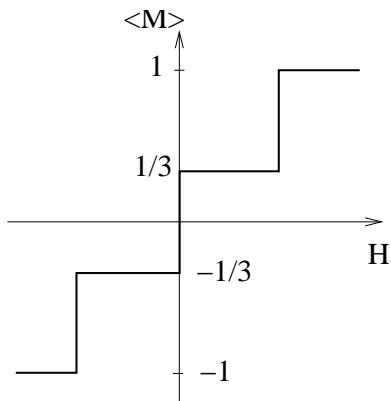}
\caption{Schematic representation of the magnetization (or filling) plateaus for a $q=0$ commensurate tube ($T=0$).}
\label{fig:q=0plateaus}
\end{figure}

Let us begin by studying the commensurate case ($q=0$) before we continue to
the more interesting incommensurate cases ($q=1,2$). As a function of the
chemical potential, the possible filling fractions are $\nu=0,1/3,2/3$, and
$1$, corresponding to the cases of no filling, one sublattice filled, two
sublattices filled, or all three sublattices filled. Notice that there is a
threefold ground state degeneracy for $\nu=1/3$ from filling any one of the
three sublattices; similarly, there is a threefold degeneracy for $\nu=2/3$
from filling all but one of the three sublattices. The only values of
chemical potential for which there is a macroscopic degeneracy in the $q=0$
case is right at the transition points between plateaus. In the spin
language, the filling fractions correspond to magnetizations $\langle M
\rangle = -1,-1/3,1/3$, and $1$. A schematic plot of the magnetization as a
function of the external field $H$ is shown in Fig.~\ref{fig:q=0plateaus}

\begin{figure}
\noindent
\center
\vspace{-.2 in}
\epsfxsize=3.2 in
\epsfbox{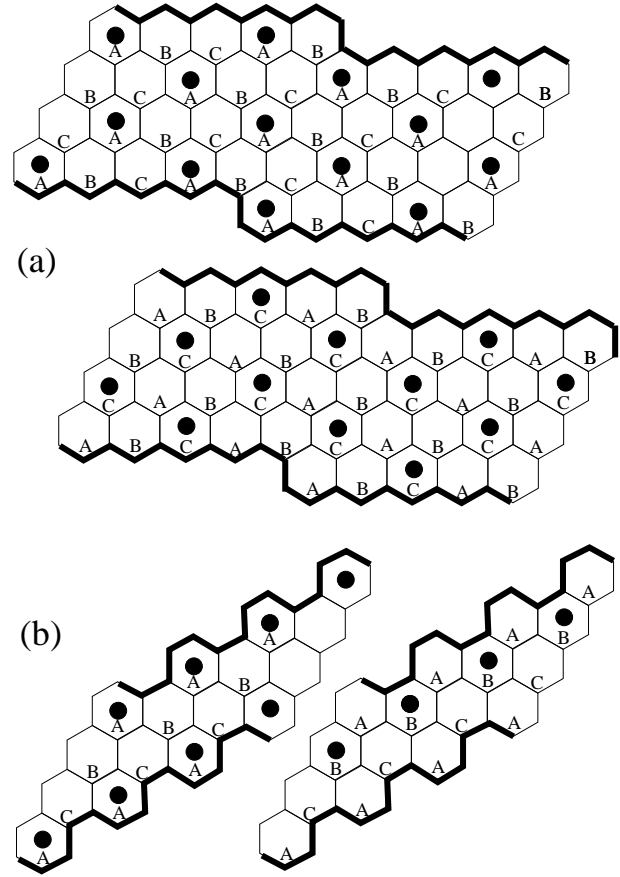}
\vspace{.3 in}
\caption{Two examples of zippers for a $(4,0)$ tube. When the tube is 
opened up along the zipper (thick lines), the hexagonal lattice can be
separated locally into three sublattices $A,B$, and $C$. By filling the
sublattices labelled by $A,B$, and $C$ one obtains for case (a) either $\nu=
7/20$ if $A$ or $B$ is chosen, or $\nu=3/10$ if $C$ is chosen. For case (b),
one obtains $\nu=1/2$ if $A$ is chosen, or $\nu=1/4$ if $B$ or $C$ is
chosen.}
\label{fig:zipper}
\end{figure}

Let us now analyze the incommensurate cases $q=1,2$. One still obtains the
trivial plateaus at $\nu=0,1$, but the question is what happens to the
partially filled plateaus. Here we describe an analytical argument that leads
to the possible filling fractions for general $(N,M)$. In this derivation, we
make some assumptions based on physical intuition, which is supported by
exact numerical transfer matrix calculations for a range of values for
$N,M$. First, one can separate the lattice sites into three classes locally,
but not globally; there will always be a topological line defect running
along the tube for $q=1,2$.  We call this line the ``zipper''; examples are shown
in Fig.~\ref{fig:zipper} for a $(4,0)$ tube. One can open up the tube along
the zipper, and label the hexagons centered at the triangular lattice points
$A,B$,and $C$. However, now the number of $A,B$, and $C$ points are no longer
all equal. Filling only one of these three classes of points leads to filling
fractions $\nu$ that depend on the zipper geometry, and on which of $A,B$, or
$C$ were chosen. For example, a case leading to filling fractions
$\nu=7/20,3/10$ is shown in Fig.~\ref{fig:zipper}a, and another leading to
filling fractions $\nu=1/2,1/4$ is shown in Fig.~\ref{fig:zipper}b.

\begin{figure}
\noindent
\center
\epsfxsize=2.0 in
\epsfbox{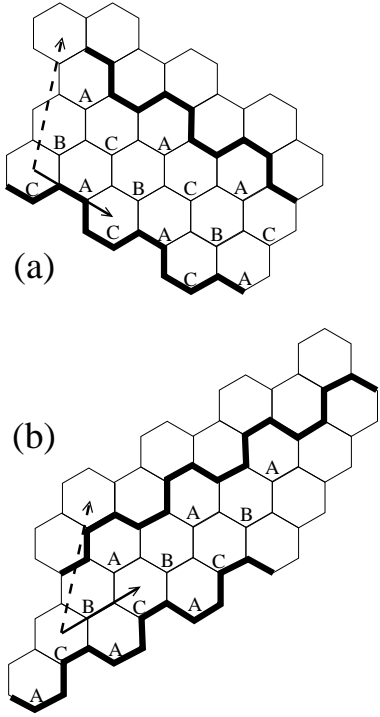}
\vspace{.1 in}
\caption{Two straight zippers in the $(2,1)$ chiral tube, labeled by
the vectors (solid arrows) $\vec z_1=(1,-2)$ in (a) and $\vec z_2=(2,-1)$ in (b). 
The wrapping vectors is the dotted arrow.}
\label{fig:two-zippers}
\end{figure}

\begin{figure}
\noindent
\center
\vspace{-.1 in}
\epsfxsize=1.8 in
\epsfbox{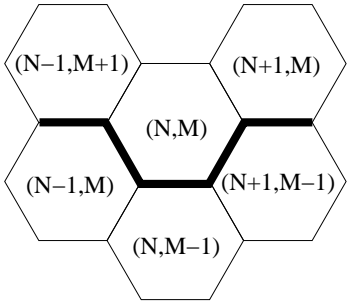}
\vspace{.1 in}
\caption{Labelling of the hexagons by their coordinates in the $\vec a_\pm$ basis for the $(2,-1)$ zipper. The
  zipper is shown as a dark line.  The region below the line lies within the boundaries of the unwrapped tube.}
\label{fig:tiling}
\end{figure}

\begin{figure}
\noindent
\center
\vspace{-.1 in}
\epsfxsize=3.0 in
\epsfbox{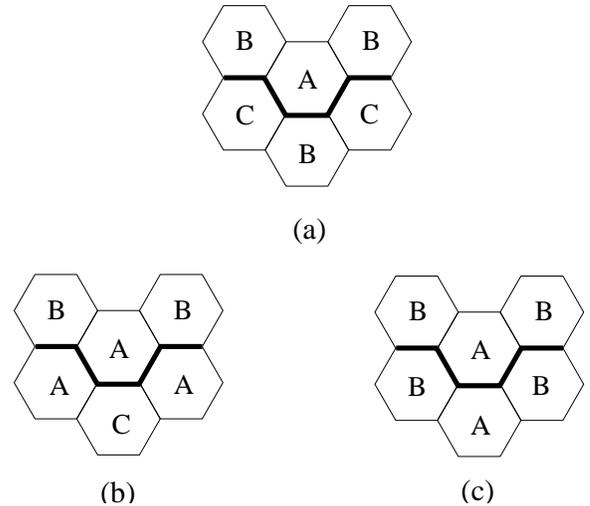}
\vspace{.2 in}
\caption{The boundary configurations for the zipper $\vec z_1=(2,-1)$ 
and the cases (a) $q=0$, (b) $q=1$, and (c) $q=2$.  The hexagons below the solid line lie within the boundaries set by the zipper and are labeled according to eqn. (\ref{eq:ij}).  Above the solid line, $(N,M)\equiv(0,0)$ is labeled by $A$, while $(N-1,M+1),(N+1,M)\equiv(1,0)$ are labeled by $B$.}
\label{fig:matches}
\end{figure}

There is an infinite number of possible zippers for a given $(N,M)$. In order
to choose the one that leads to the lowest energy state, we argue that the
zipper must be as straight as possible. The rationale is that bending the
zipper should cost energy. So zippers as in Fig.~\ref{fig:zipper}b should be
preferred. For a given wrapping vector $(N,M)$, $N,M>0$, we then investigate
two zippers that we label by two vectors $\vec z_1=(1,-2)$ and $\vec z_2=(2,-1)$. These
two vectors differ by a rotation of 60$^\circ$, and are shown in
Fig~\ref{fig:two-zippers}. A third $\vec z_3=(1,1)$ is not considered because for
$N,M>0$ this zipper winds rapidly around the tube, and hence there is a
large energy cost associated with its longer length. 
%
To proceed, we must calculate the boundary energy cost associated with the
topological mismatch on the two sides of the zipper. Consider the
portion of a tube shown in Fig~\ref{fig:tiling}, for a $(2,-1)$ zipper, where in
each hexagon we write its coordinates $(i,j)$ in the ${\vec a}_\pm$ basis.
Within the boundaries set by the zipper, we can label the sites $A,B$, and $C$ according to: 
\begin{equation}
i-j\equiv \cases{
0\; ({\rm mod}\; 3) \to A \cr
1\; ({\rm mod}\; 3) \to B \cr
2\; ({\rm mod}\; 3) \to C }\; .
\label{eq:ij}
\end{equation}
In Fig.~\ref{fig:matches} we show the possible boundary configurations for
$q=0,1,2$.  Let $n_b$ be the number of near neighbors that are now both occupied
{\it per} unit length, $|\vec z_2|$, along the zipper. (We also refer to the situation of
occupied near neighbor sites as a {\it bond} between the sites). For $q=0$, $n_b=0$ since the tripartite lattice is not frustrated, but
\begin{itemize}
\item{for $q=1$, $n_b=0$ if $B$ or $C$ is the filled (minority)
    sublattice; $n_b=2$ if $A$ is the filled (majority) sublattice.  }
\item{for $q=2$, $n_b=0$ if $C$ is the filled (minority) sublattice;
    $n_b=1$ if $A$ or $B$ is the filled (majority) sublattice.  }
\end{itemize}
Each set of occupied neighbors costs energy $V$. The energy $E(\nu)$ {\it per}
area ${\cal A}$ for a given filling configuration is
\begin{equation}
\frac{E(\nu)}{\cal A}= \frac{n_b V}{\cal A}- \nu \mu\;,
\end{equation}
where ${\cal A}$ is the supercell area given by 
\bea
\label{eq:supercellA}
{\cal A}&=&(N{\vec a}_+ + M{\vec a}_-)\cdot\vec z_2\non\\
&=&(N{\vec a}_+ + M{\vec a}_-)\cdot(2 {\vec a}_+ - {\vec a}_-)\non\\
&=& (2M+N) {\cal A}_{\rm hex}
\eea
and ${\cal A}_{\rm hex}=|{\vec a_\pm}|^2\sqrt{3}/2$ is the area of one hexagon,
which we will set to unity.
Given that the number of $A,B$, and $C$ lattice points are not all equal for
$q\ne 0$, the possible filling fractions depend on the choice of filling
either majority or minority lattice points. It is simple to check that
\begin{eqnarray}
\nu_+&&=\frac{\lceil {\cal A}/3\rceil}{{\cal A}} \quad 
\hbox{for majority filling}\\
\nu_-&&=\frac{\lfloor {\cal A}/3\rfloor}{{\cal A}} \quad 
\hbox{for minority filling}
\end{eqnarray}
These fillings can be converted to magnetization values by $m_\pm=\langle M
\rangle= 2\nu_\mp-1$ and $H=\mu-3V$. Notice that the spin flip symmetry $m \to -
m$ corresponds to the particle-hole symmetry $\nu \to 1- \nu$.

From this point on, the analysis is straightforward; one has to consider the
energy density ${E(\nu)}/{\cal A}$ for {\it all} the different possibilities
$q=1,2$, zippers $\vec z_{1,2}$, as well as majority or minority fillings
$\nu_\pm$.  The transition between $\nu_+$ and $\nu_-$ occurs at a critical field $H_c$ when $E(\nu_+)=E(\nu_-)$.  After considering the eight cases, we obtain the minimum energy configurations for a generic $(N,M)$ tube with $q=1$:
\begin{eqnarray}
&\;& m_+=\frac{1}{3}\left(1+\frac{2}{2M+N}\right) \quad
m_-=\frac{1}{3}\left(1-\frac{2}{2N+M}\right) \nonumber\\
&\;&\hspace{2cm}
H_c=\left(4-\frac{2M}{N+M}\right)V
\label{eq:Exact}
\end{eqnarray}
The complementary case of $q=2$ is obtained by interchanging
$N\leftrightarrow M$.  We have verified the magnetizations and critical
fields of eqn. (\ref{eq:Exact}) for small values of $N,M$ previously
\cite{Green-Chamon}.

Notice that we recover the graphite (planar) result $m_\pm=\pm1/3$ as the
tube radius approaches infinity ($N,M\rightarrow\infty$), as we should.

\subsection{Macroscopic Degeneracy}
\label{sec:degeneracy}

Most plateaus for general $(N,M)$ tubes have only a discrete degeneracy
corresponding to filling different sublattices.  There is a macroscopic
degeneracy only at the transition points.  However, a careful consideration
of the plateaus in the exceptional case of the frustrated zig-zag $(N,0)$
tubes reveals that either the $m_+$ or $m_-$ plateau is macroscopically
degenerate for its whole range of stability as a function of chemical
potential (or magnetic field). Which plateau is degenerate depends on the the
value of $q$.

\begin{figure}
\epsfxsize=3.25in
\epsfbox{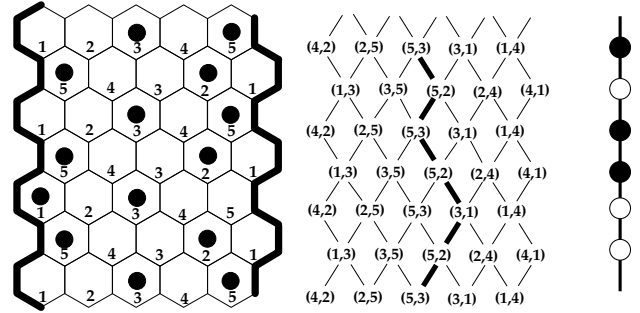}
\vspace{.3in}
\caption{LEFT: Typical configuration in $\nu_+$ (or $m_-$) of the
$(5,0)$ tube.  Alternating numbering within layers allows a symmetric
description from bottom-to-top or top-to-bottom.  CENTER: Allowed states as
paths on a wrapped square lattice.  The vertex labels may be dropped. RIGHT:
Steps to the right can be represented by particles (solid circles), and to
the left by holes (open circles).}
\label{Hop-Lattice}
\end{figure}

Let us reconsider the $\nu_+$ filling of the $(5,0)$ tube in
Fig.~\ref{Hop-Lattice}.
Imagine building a typical $\nu_+$ state layer-by-layer from top to bottom,
with a total of $L$ layers.  Each new layer must add exactly two filled sites
and one nearest-neighbor pair of occupied sites (1 bond). There are two ways
to achieve this constraint. A non-trivial one is such that no two adjacent
sites may be occupied within a layer, and this is the case shown in
Fig.~\ref{Hop-Lattice} (left). Notice that moving a particle from a layer to another
adds another intra-layer bond. There is also a trivial one where two adjacent
sites are occupied within a layer, and to conserve $n_b$, two adjacent sites
must be occupied in the next, and so on down the tube. However, the class of
such trivial states is only 5-fold degenerate, as opposed to the
macroscopically degenerate class containing the state shown in
Fig.~\ref{Hop-Lattice}. We now turn to the enumeration of the non-trivial
class of states.
An allowed state can be represented as a string of occupied sites,
$\sigma=\{\sigma_i\}$, $i=1,\ldots,L$, which in our example is
$\sigma=\{\cdots(5,3)(5,2)(5,3)\cdots\}$.  At each layer, there are exactly
two possibilities for the following one.  For example, $(1,4)$ can be
followed by $(1,4)$ or by $(2,4)$.  However, the total number of
possibilities at any given level is five.  Fig.~\ref{Hop-Lattice} (center)
summarizes this structure succinctly as a {\em square} lattice wrapped on the
cylinder.  A typical state, then, is a lattice path along the tube.  A path
on this type of lattice is known as a Bratteli diagram.  There is a recent
Hubbard model considered by Henley and Zhang \cite{Henley} of spinless
fermions on a square lattice in which the bookkeeping of states is very
similar.

Generalizing to $(N,0)$, we find $N$ possible states in each layer and two in
the succeeding one, and the structure of states is again that of a wrapped
Bratteli diagram with $N$ sites along the circumference.  The dimension of
the Hilbert space is the number of lattice paths, $N2^L$, so that in an
infinitely long tube, the entropy per site (in the thermodynamic $L\to
\infty$ limit) is exactly $S=({\rm ln}2)/N$.  Notice that constrained paths
introduce correlations along the length of the tube, despite the absence of
inter-layer hopping.

The preceeding discussion is valid for all $N$ with $q=2$, where the
macroscopically degenerate plateau occurs at $\nu_+$.  For $q=1$, the only
modification is that the macroscopically degenerate plateau has filling
$\nu_-$.

\section{Quantum limit}
\label{sec:quantum}

Having understood in detail the classical limit, we now turn on a small
hopping, $t\ll V$, that introduces quantum fluctuations.  Away from the
transitions, adding a particle necessarily increases $n_b$ so that all
plateaus begin with a classical gap of order $V$.  Consequently, we can
project to the Hilbert space of the classical ground states. Those plateaus
which have only a discrete symmetry must retain their gaps, but the
macroscopically degenerate plateaus are more complicated.

The matrix elements of the projected Hamiltonian (\ref{eq:Hubbard}) connect
only those states that differ by a single hop of a boson:
\begin{equation}
\langle\sigma^\prime|{\cal H}|\sigma\rangle = 
\cases{
-2t\quad{\rm if}\quad\sum_i\delta_{\sigma_i \sigma^\prime_i} = L-1\cr
0~~~{\rm otherwise,} }
\label{eq:Hopping}
\end{equation}
where $\sigma,\sigma^\prime$ are two paths in the Bratteli lattice constructed in
Section \ref{sec:degeneracy}.

It turns out that this Hamiltonian is exactly solvable, being closely related
to a class of solid-on-solid models that were introduced by Pasquier
\cite{SOS}.  In the following, we derive the continuum limit of ${\cal H}$ directly.

In the previous section, the paths $\sigma$ were labeled, for clarity, by a
string of occupied sites on the nanotube.  A simpler representation is to
work with the wrapped square lattice directly, where the path is uniquely
specified by an initial point and its direction in each layer.  We will label
the topmost layer by $i=1$ with $i$ increasing by $1$ with each downward
move.  There are $L$ layers of hexagons and we impose periodic boundary
conditions, $L+1\equiv1$.  In order for the layers to match, $L$ has to be
even.

To specify the initial point on $\sigma$, we chose an ``anchor'' $\alpha$ on
one of the $N$ sites in the $i=1$ layer ($\alpha=1,\ldots,N$).  Now, we
represent a step to the right in layer $i$ by a fermion, and a step to the
left by a hole. We define fermion creation and anihilation operators
$c^\dagger_i$ and $c_i$ for each layer $i$. A state $|\sigma\rangle$ in the
Hilbert space, $S$, is represented by
\bea |\sigma\rangle=|\alpha\rangle\otimes
c^\dagger_{i_1}c^\dagger_{i_2}\cdots c^\dagger_{i_p}|0\rangle~,
\label{eq:state}
\eea 
where $\alpha$ is the anchor site in the first layer and $i_1\cdots i_p$ are
the layers where the path steps to the right.  For instance, the portion of
the path in Fig. \ref{Hop-Lattice} is $|\sigma\rangle=|3\rangle\otimes
c^\dagger_1c^\dagger_3c^\dagger_4\cdots |0\rangle$.  The fermionic
representation is convenient since there is exactly one step in each layer,
but hard-core bosons can also be used, as in one dimension they are
equivalent. The number of particles (steps to the right) and holes (steps to
the left) must add up to $L$ (the total number of steps). Each path also has
a topological character for the number of times that it winds around the
tube, which must be a multiple of $N$ in order for the path to close with the
periodic boundary condition $L+1\equiv 1$. These two requirements are summarized by
\bea
N_p+N_h&=&L\\
N_p-N_h&=&bN~,
\label{eq:charge}
\eea    
where $N_{p,h}$ is the number of particles or holes, and $b$ is the winding number (positive or negative).  If each particle (hole) is assigned a positive (negative) unit charge, then $bN$ is the total charge. Notice
that $b=0$ at half-filling ($N_p=N_h=L/2$).

Whenever it is allowed within a layer, a single hop changes the step sequence
right-left to left-right and {\it vice versa}, which corresponds to
$c^\dagger_{i+1}c_i$ or $c^\dagger_ic_{i+1}$.  In a layer without a kink no
hops are possible, and the hopping terms vanish by fermionic statistics.
Since we are working in periodic boundary conditions, the boundary terms,
$c^\dagger_1c_L$ and $c^\dagger_Lc_1$, must be treated more carefully.  A hop
at this point is necessarily accompanied by a translation of the anchor point
by $|\alpha\rangle\mapsto|\alpha\pm 1\rangle$.  Let us represent this
operation by \bea R_\pm|\alpha\rangle=|\alpha\pm 1\rangle~ \eea with
$R^\dagger_-=R_+$.  Cylindrical wrapping requires a ${\bf Z}_N$ symmetry
because ${|\alpha\pm N\rangle\equiv|\alpha\rangle}$, {\it i.e.} $R_\pm^N=R_\pm$.
Putting the bulk and boundary hopping terms together, the Hamiltonian of eqn.
(\ref{eq:Hopping}) becomes \bea {\cal H} =
-2t\left[\sum_{i=1}^{L-1}c^\dagger_{i+1}c_i+R_-\otimes c^\dagger_1c_L\right]
+ h.c.~.
\label{H_hopping}
\eea 
We can think of ${\cal H}$ as describing free fermions on a periodic one
dimensional chain with a ${\bf Z}_N$ impurity on one of the bonds.

${\cal H}$ can be diagonalized exactly in momentum space.  Going around the
tube lengthwise contributes a phase $e^{ikL}$ while going around the
perimeter contributes $e^{i\phi}$, with $\phi = 2\pi a/N$ ($a=1,\ldots,N-1$).
Therefore toroidal boundary conditions require
\bea 
\label{eq:phases}
e^{ikL}e^{i\phi} = 1~.
\eea 
Or, 
\bea k=\frac{2\pi v}{L}\left(n+\frac{a}{N}\right)~,
\label{eq:k}
\eea
where $v=2t$ is the velocity and $n$ is an integer.  The Hamiltonian contains
the usual free particle dispersion,
\bea
\label{eq:H_fermions}
{\cal H}=-4t\sum_k \cos k\,c^\dagger_k c_k~,
\eea
but the allowed $k$ are given by eqn. (\ref{eq:k}).
If the spectrum (near half-filling) is linearized around the Fermi momentum,
$|k_F|$, then at small $k$ states with nonzero $a$ cost an additional energy
of $(2\pi v/L)(a/N)^2$.
 
\begin{figure}
\epsfxsize=3.0in
\epsfbox{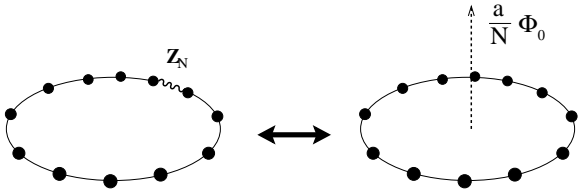}
\vspace{.4 in}
\caption{The left ring shows the ${\bf Z}_N$ impurity on the anchor bond (wavy
  line).  The right ring shows the equivalent alternative, where the impurity
  is replaced by a flux tube through the torus.}
\label{fig:flux}
\end{figure}

The $a/N$ offset in $k$ can be thought of as a minimally coupled vector
potential such that the magnetic field is a $\delta$-flux tube through the
torus containing $a/N$ flux quanta.  In other words, one of the bonds along
the chain (the anchor) had ${\bf Z}_N$ symmetry, whose effect is equivalent
to a flux tube.  Fig. \ref{fig:flux} illustrates this equivalence.
The offset in $k$ due to the flux induces a total current in the fermion
system.

At this point, one can see two topological effects on the torus.  The first
is the ${\bf Z}_N$ flux tube, or total current.  As we have seen, its
contribution to the energy near $|k_F|$ was $(2\pi v/L)(a/N)^2$.  The second
is the path winding along the length of the tube, or the total charge defined by eqn. (\ref{eq:charge}).  Its contribution to the energy near $|k_F|$ is
similar, $(2\pi v/L)(bN/2)^2$. The total energy due to these topological
sectors is
\bea
\label{eq:E_top}
\Delta E_{a,b}=\frac{2\pi v}{L} 
\left(\frac{a^2}{N^2}+\frac{b^2 N^2}{4}\right)\,.
\eea 
This expression is familiar from the Luttinger liquid model of
one-dimensional spinless Fermions \cite{Haldane}.

We can now obtain the continuum limit of our model.  It is well known that
free fermions in one dimension are equivalent to free bosons.  The
corresponding Lagrangian is
\bea 
{\cal L} =\frac{1}{8\pi}[v^{-1}(\partial_t\varphi)^2 -
v(\partial_x\varphi)^2]~, 
\eea
where $\varphi$ is the bosonic field.  ${\cal L}$ is a conformally invariant
theory with central charge $c=1$.  We conjecture that the topological effects
that we described above come from compactifying $\varphi$ on a circle of
radius $R$,
\bea 
\phi\equiv\phi+2\pi R~.
\label{eq:compactification}
\eea 
By compactifying the boson, topological modes (or zero modes) appear.  In
field theory, they are conventionally obtained from electric and magnetic
monopoles.  The energy of the zero modes is
\bea
E^0_{a,b}=\frac{2\pi v}{L} \left(\frac{a^2}{R^2}+\frac{b^2 R^2}{4}\right) ,
\label{eq:ZeroModes}
\eea 
where $a$ and $b$ are integers labeling the fundamental cycles on the torus.
Comparing $E^0_{a,b}$ (\ref{eq:ZeroModes}) to $\Delta E_{a,b}$ (\ref{eq:E_top}), we find that
$R=N$.  The ordinary phonon (oscillator) modes exist on top of each
topological sector and simply contribute the usual phonon energy, so that the
complete dispersion is
\bea
E=E^0_{a,b}+\frac{2\pi}{L}|n|~.
\eea 
We have verified this energy spectrum by exact diagonalization for small
systems in an earlier work \cite{Green-Chamon}.

The overall picture of a compactified boson with central charge $c=1$ is
consistent with the solid-on-solid models of Pasquier \cite{SOS}.  One can
also ask what happens at higher order in $t/V$, which is the subject of the
next subsection.

\subsection{Finite $t/V$}
\label{eq:finite_t}

The next corrections to ${\cal H}$ are of order $t^2/V$ and involve virtual transitions to
adatom configurations that are not in the degenerate subspace $S$.  The
generic form is
\bea -\frac{t^2}{V}\,{\cal P}_S\left[\sum_{{\langle
ij\rangle}{\langle kl\rangle}}b^\dagger_ib_jb^\dagger_kb_l\right]{\cal P}_S~,
\eea 
where ${\cal P}_S$ is a projection operator into $S$.  Another way of writing the second order perturbation is the familiar form,
\bea
\langle\sigma^\prime|{\cal H}|\sigma\rangle \rightarrow 
\langle\sigma^\prime|{\cal H}|\sigma\rangle - \sum_\lambda\frac{\langle\sigma^\prime|{\cal H}|\lambda\rangle\langle\lambda|{\cal H}|\sigma\rangle}
{E_\lambda-E_\sigma}~,
\label{eq:virtual}
\eea
where $|\sigma\rangle, |\sigma^\prime\rangle\in S$ while $|\lambda\rangle\notin S$ is the intermediate virtual state with energy $E_\lambda$.  $E_\sigma$ and $E_{\sigma^\prime}$ are, of course, equal and constant, and all energy differences are due to the nearest neighbor repulsion $Vn_in_j$. 

There are three types of virtual processes: (i) single particle hopping from $\sigma$ to $\sigma\neq\sigma^\prime$, (ii) two particle correlated hopping from $\sigma$ to $\sigma\neq\sigma^\prime$ and (iii) single particle diagonal hopping from  $\sigma$ back into $\sigma$.  For concreteness, consider process (iii) in the $(5,0)$ state of fig. \ref{Hop-Lattice}.  Whenever there is a kink in $\sigma$, such as in the third layer from the top, the contribution to eqn. (\ref{eq:virtual}) from all virtual hops is $-(35/6)4t^2/V$.  For example, the adatom on site $3$ can hop into any one of its six neighbors with the energy denominators $1/2+1/2+1/2+1/2+1/2+1/3$ (in units of $t^2/V$).  Similarly, the adatom on site $5$ contributes $1+1+1$, for a total of $35/6$ (it is forbidden to hop one site over to the right because the resulting state is in $S$).  On the other hand, if there is no kink, the contribution is $-8\cdot 4t^2/V$.  The criterion for a kink in layer $i$ is $2[1/4-(\w n_i-1/2)(\w n_{i+1}-1/2)]=1$, where $\w n_i=c^\dagger_ic_i$ is the fermion number; otherwise this quantity vanishes.  Similarly, the absence of a kink is synonymous with the nonvanishing of $2[1/4+(\w n_i-1/2)(\w n_{i+1}-1/2)]$.  Thus, the total diagonal contribution to ${\cal H}$ can be written
\breakon
\bea
\label{eq:4fermion}
{\cal H}&&\rightarrow{\cal H}-
\frac{8t^2}{V}\sum_\sigma\sum_i\left\{\frac{35}{6}
\left[\frac{1}{4}\!-\!\left(\w n_i\!-\!\frac{1}{2}\right)
\left(\w n_{i+1}\!-\!\frac{1}{2}\right)\right]\!+\!
8\left[\left(\w n_i\!-\!\frac{1}{2}\right)\left(\w n_{i+1}\!-
\!\frac{1}{2}\right)\!+\!\frac{1}{4}\right]\right\}|\sigma\rangle\langle\sigma|
\non\\
&&={\cal H}-\frac{4t^2}{V}\sum_\sigma\left[\frac{13}{3}\sum_i \w n_i \w n_{i+1}-
\frac{13}{3}\sum_i \w n_i +8\right]|\sigma\rangle\langle\sigma|~.
\eea   
\breakoff

For general $N$, the correction scales like $N$. The essential term in the last line of eqn. (\ref{eq:4fermion}) is the first one.  This four-fermion interaction renormalizes the radius $R$ by corrections of order $t/V$.

Let us return to processes (i) and (ii).  An example of (i) is the adatom in the third layer from the top, site $5$, hopping to the second layer, site $1$, and then back into the third layer, site $1$.  The intermediate state is not in $S$.  This process serves only to renormalize $t$ because its amplitude is the same for all kinks.  An example of (ii) is the particle in the fourth layer, site $5$, hopping to site $4$, {\it followed by} the particle in the third layer, site $5$, hopping to site $1$ in the same layer.  This correlated hopping occurs in a configuration containing the sequence particle-hole-hole or hole-particle-particle, which corresponds to a next-nearest neighbor interaction $c^\dagger_{i+2}c^\dagger_{i+1}c_{i+1}c_i+h.c.$.  This type of term has a similar effect to the $\w n_i\w n_{i+1}$ term in eqn. (\ref{eq:4fermion}), it reduces the value of $R$ by corrections of order $t/V$.  The four-fermion terms do not cancel unless there is a delicate cancellation.   
\begin{figure}
\epsfxsize=2.2in
\center
\epsfbox{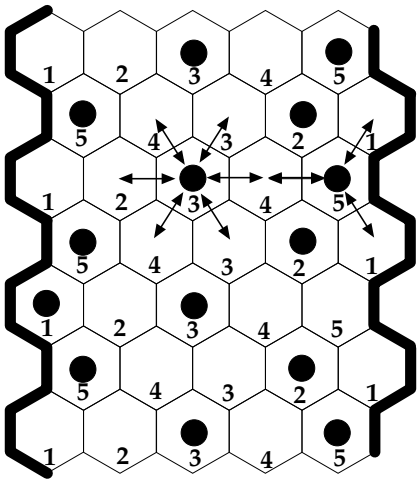}
\vspace{.2in}
\caption{An example of the virtual process (iii) in the configuration of
Fig. \ref{Hop-Lattice}.}
\label{fig:virtual}
\end{figure}
One can also consider the extreme limit in which $t\gg V$.  In this case, the
XXZ Hamiltonian in eqn. (\ref{eq:Spin}) is simply the XY model on a cylinder.
Note that our model has the particle-hole symmetry $t\leftrightarrow -t$ at
least to order $t^2/V$.  Let us denote the spin angle relative to the
cylindrical surface by $\varphi(x,\theta)$, where $x$ is the coordinate along
the tube and $\theta$ is the coordinate around the perimeter.  Uniqueness of
the wavefunction requires that $\varphi$ has the periodicity
$\varphi(x,\theta+2\pi)=\varphi(x,\theta)+2\pi m$, where $m$ is an integer.
The low energy excitations are purely along the length of the tube;
excitations around the perimeter will cost an energy on the order of $1/N$,
which is large compared to $1/L$.  Thus we can freeze the $\theta$
coordinate, and the energy density is proportional to
$|\partial_x\varphi|^2$.  Since the periodicity is still $\varphi\equiv
\varphi+2\pi m$, we end up with a free boson compactified on radius
$R_{_{\rm XY}}=1$.  The question is how the adsorption regime $t\ll V$, which is
also a compactified boson but on radius $R=N$, is reached.

\subsection{The Lieb-Schultz-Mattis Argument}
\label{sec:LSM}

Another indication that our plateaus are nontrivial comes from trying to
understand the spin tube as a spin ladder.  Let us take this point of view to
see if it yields the zero gap.  A standard approach is to use a
Lieb-Schultz-Mattis (LSM) argument, in which the spins are deformed slowly
along the length\cite{Oshikawa}.  Applying it to our tube, we find that a
plateau is gapless if $\hat S-\hat M$ {\it is not} an integer, where $\hat S$ and $\hat M$ are
the total spin and magnetization, respectively, per layer (a layer being the
$N$ sites around the perimeter).  Using $\hat S=N/2$ and the magnetizations from
Eqn.~(\ref{eq:Exact}), we find that $\hat S-\hat M$ {\it is} an integer in the
macroscopically degenerate plateaus, so that the LSM argument is insufficient
in this case. Therefore, a conclusive argument must take the geometric
frustration into account.

\subsection{Special Case: $N=2$}
\label{sec:N=2}

\begin{figure}
\epsfxsize=3.0in
\epsfbox{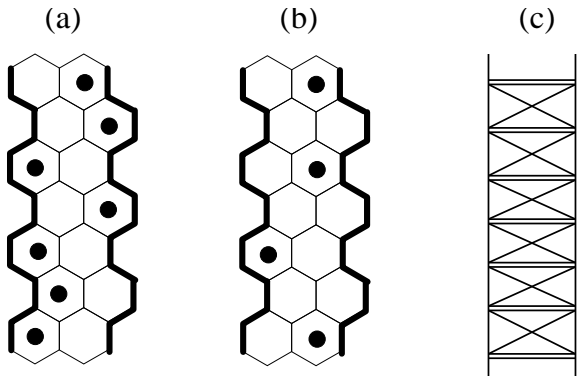}
\caption{The fillings for $N=2$. $\nu_+=1/2$ (a) and $\nu_-=1/4$ (b). 
  Each adatom can live at either site in its layer because each site is
  connected to every site in the neighboring layers.  (c) shows the
  triangular lattice in the plane; the horizontal double bond is due to the
  periodicity around a cylinder.  (c) is exactly the geometry of the spin
  ladder studied by other authors (albeit with different coupling).}
\label{fig:N=2}
\end{figure}

Before concluding with the effective theory, we should point out that the
geometry of the $(2,0)$ tube is special; all sites in adjacent layers are
interconnected.  As a result, all of its plateaus have an extensive entropy,
and we find that hopping opens a gap in both plateaus.  Fig. \ref{fig:N=2}
illustrates this exception.  At either filling, the adatom in each layer is
free to hop to either site---both configurations are iso-energetic because
each site is contiguous to all sites in the neighboring layers.  Hence both
plateaus are macroscopically degenerate.  In the presence of hopping, each
adatom lives in a double well potential, which has a gap of order $t$.
Therefore, neither plateau is gapless unlike higher $N$.

In fact, this tube can be written as a spin chain that has been studied at
isotropic coupling \cite{Sakai}, $-2t=V$.  Two plateaus were found in this
case, and it is tempting to speculate whether the two regimes are connected
adiabatically.
Another way to understand the gap is within the fermion model.  Due to the
high connectivity of the Bratteli diagram for $N=2$, there is another allowed
hop in addition to the $c^\dagger_ic_{i+1}+c^\dagger_{i+1}c_i$ term, which is
$c^\dagger_ic^\dagger_{i+1}+c_{i+1}c_i$.  This is a pairing interaction that
leads to a superconducting gap.

\section{Discussion and Conclusion}
\label{sec:conclusion}

In this paper we studied spin and Bose-Hubbard models in tube geometries. We
found that an undelying frustrated triangular lattice, combined with extra
geometric frustration from the closed topology of the tube, leads to
interesting plateau structures for filling fractions $\nu$ (in the
Bose-Hubbard models) or magnetizations $\langle M \rangle$ (for the spin
models). In section \ref{sec:classical} we studied the classical limit when
the hopping $t$ in the Bose-Hubbard model goes to zero; in the spin language,
this correspond to the Ising limit. We obtained the different filling
fractions allowed energetically by the wrapping condition, labelled by the
vector $(N,M)$. We showed that the corresponding solid phases are unique
states, with the exception of the zig-zag tubes $(N,0)$ with $N$ not
divisible by 3. In this particular case, which we studied in section
\ref{sec:degeneracy}, we showed that there is a macroscopic degeneracy of the
classical ground state for a range of chemical potentials (or fields), {\it
i.e.} a degenerate plateau. In contrast, when $N \equiv 0\; ({\rm mod}\;
3)$, there is only a single value of the chemical potential (or magnetic
field, $H=0$) for which the ground state degeneracy is non-trivial. We
enumerated all the degenerate states at the plateau by mapping each state in
the degenerate Hilbert space into a path in a Bratteli diagram wrapped on a
cylinder. Using this enumeration, we then turned on the quantum hopping term
$t$ (or the XY spin couplings $J_\perp=2t$) in section
\ref{sec:quantum}. We found that the quantum terms lift up the degeneracy,
leaving a gapless spectrum described by a $c=1$ conformal field theory with
compactification radius $R=N$. We presented an analytical argument that shows
that the Bratteli path can be mapped into a state of a
fermion model in a one dimensional chain, and the hopping terms in the tube
lattice correspond to a hopping in the fermion chain. The compactification
radius follows from a boundary ${\bf Z}_N$ degree of freedom and path
winding, which must be included when the system is subject to periodic
boundary conditions. We also discussed the cases when the anisotropy $t/V$ (or
$J_\perp/J_z$) is no longer small, and the special case $(N=2,0)$.

There are a number of interesting features that emerge from this problem of
restricting spin and Bose-Hubbard Hamiltonians to tube geometries. For
example, it is noteworthy that in the case of the $(N,0)$ zig-zag tubes with
$N$ not divisble by 3, the low energy spectrum in the quantum case is gapless
and described by a conformal field theory with a quantized compactification
radius $R=N$; in the language of Luttinger liquids, this means that the
Luttinger parameter is fixed by topology, similarly to the case of edge
states in a fractional quantum Hall fluid\cite{Wen}, and in contrast to
quantum wires (where the Luttinger parameter can vary continuously). In the
spin tube problem, this quantization, in the projective limit $J_\perp \ll
|J_z|$ (or $t \ll V$), follows simply from the fact that there is a direct
connection between the physical lattice and the target space for the bosons.
One of the interesting features of the spectrum for the gapless plateaus is
that the specific heat is linear in temperature $T$, with the prefactor
related to the velocity and a universal constant that depends on the central
charge $c=1$: $C=c\frac{\pi vk_B^2}{3}T=\frac{\pi vk_B^2}{3}T$. This
dependence would be manifest in thermal measurements, and could be contrasted
with exponentially activated behavior for the gapped solid phases.

Recently, there has been some interest in a connection that was pointed out
by Kitaev \cite{Kitaev} between the topological stability in some class of spin
Hamiltonians in toroidal structures and quantum error correction for quantum
computation. Bonesteel \cite{Bonesteel} investigated the possibility of using
the two-fold topological degeneracy of spin-1/2 chiral spin liquid states on
the torus to construct quantum error correcting codes. Since closed carbon
nanotube structures (tori) have been observed experimentaly \cite{nature}, it
may be possible that spin tube geometries such as the ones discussed in this
paper could realize physically some related Hamiltonians. We note, however,
that for the class of Hamiltonians with nearest neighbor couplings that we
studied here, there was only a trivial non-degenerate ground state, although
there were topologically non-trivial excitations.

Finally, we would like to discuss a possible experimental set up for
observing the filling fraction plateaus in the case of the monolayer
adsorption problem of rare gases on the surface of carbon nanotubes. The
filling fraction plateaus could be probed by measuring the resonance
frequencies for a vibrating single wall carbon nanotube as a function of the
vapor pressure of rare gas in a chamber. One way of measuring the resonant
frequencies would be to have a single-wall tube hang alongside a
multiwall tube, both with metallic grains at their tips. The multiwall tube
is very stiff, and is basically rigid as compared to the single wall. By
measuring the changes in the capacitance between the grains, one could probe
the frequency of vibrations of the single-wall tube. This frequency is given
by the square root of the ratio between the linear mass density and the
stifness of the tube. Different filling fractions of adsorbed noble gases
change the mass density of the tubes; however, their effect on the tube
stiffness should be insignificant. The change in mass density of the nanotube
due to adsorption of helium is basically $\delta \rho/\rho=\nu \ M_{\rm
He}/2M_{\rm C}=\nu/12$. Hence, the ratio between the resonant frequencies
$\omega_{1,2}$ for two plateaus at $\nu_{1,2}$ is
\begin{equation}
\frac{\omega_1}{\omega_2}=
\sqrt{\frac{1+{\nu_1 \over 12}}{1+{\nu_2\over 12}}}
\approx 1+\frac{\nu_1-\nu_2}{24}
\; .
\end{equation}
Resonant frequencies can be measured very precisely, so the filling fraction
steps $\nu_1-\nu_2$ should be experimentally measurable as a function of the
vapor pressure of helium.

\vspace{.5in}
\acknowledgements{
The authors wish to thank C.~Henley for interesting discussions, and N.~Read
for helpful comments and for pointing out Pasquier's work and the terminology
of the Bratteli diagram. Support was provided by the NSF Grant DMR-98-18259
 (D.~G.), DMR-98-76208 and the Alfred P. Sloan Foundation (C.~C.).  }

\end{multicols} 

\end{document}